\documentclass[12pt]{article}
\usepackage{amssymb}

\begin{document}

\begin{center}
\bigskip

\bigskip {\large \textbf{Metrics Fluctuational Theory}}

\medskip Timur.~F.~Kamalov

Physics Department, Moscow State Opened University,

ul. P. Korchagina, 22, Moscow 107996, Russia

Tel./fax 7-095-2821444

\begin{tabular}{ll}
E-mail: & okamalov@chat.ru \\ 
& ykamalov@rambler.ru
\end{tabular}
\end{center}

It is supposed the alternative to Quantum Mechanics Axiomatic. Fluctuational
Theory save the Mathematics of Quantum Mechanic without change, naming this
Mathematics as Method of Indirect Computation.

Fluctuational Theory is delete the axiomatic of Quantum Mechanics and
replaces it by the assumption of Gravitational Noise. This assumption is
connects the Method of Indirect Computation to the Classical Physics.

Physical fluctuations of classical gravitational fields are mathematically
expressed through geometric fluctuations of metrics of Riemann Space.

Metrics Fluctuational Theory and Quantum Mechanic is describe the classical
experiment of electrons interference by two different way.

Pacs 03.65*

Key words: Fluctuational Theory, Gravity Background, Gravity Noise,
Polarization Variables, Entanglement State.

\begin{center}
Introduction
\end{center}

Classical particles in stochastic classical gravitational fields called here
the microobjects in random gravitational background. The latter is assumed
to be the background of stochastic gravitational fields and waves
distributed isotropically over the space-time in the average. The concept of
probability interval in the Riemann space is introduced. The interference
experiment with electrons is described here. Physical fluctuations of
classical gravitational fields are mathematically expressed through
geometric fluctuations of metrics of 4-dimensional Riemann space.

\begin{center}
Main part
\end{center}

Quantum mechanics is the theory with axiomatic is based on basic results
obtained in a number of experiment studies. These experiments were
impossible to explain within the framework of classical physics, and they
constituted the foundations of the new theory called quantum one.

Though this new theory did not investigate the nature of experiments laid in
its foundation, as neither theory would investigate its own axioms,
nevertheless, it enabled performing specific calculations of real physical
experiments and was capable of correctly predicting the behavior of quantum
micro-objects. Let us start considering these experiments from the viewpoint
of classical physics, and further pass to their discussion with the help of
quantum theory.

Let us agree to consider the physical gravitational field really existing,
its approximate mathematical description to be expressed through geometrical
fluctuations of spatial metrics. This spares us difficulties with
energy-momentum conservation, as we consider the physical field to transfer
energy - in contrast with the mathematical description of gravitational
effects with Riemann geometry where the geometry is usually considered not
to contain and not to transfer energy-momentum. In this, we shall confine
ourselves to the linear approximation of metrics decomposition, which
describes rather comprehensively all known experiments in verification of
the General Relativity Theory.

Let us consider two classic particles in a field of random gravitational
fields or waves. The General Theory of Relativity gives the length element
in 4-dimensional Riemann space as

\begin{center}
$\Delta \ell ^{2}=g_{ik}\Delta x^{i}\Delta x^{k}$,
\end{center}

the metric in the linear approach is

\begin{center}
$g_{ik}=\eta _{ik}+h_{ik}$,
\end{center}

being $\eta _{ik}$ Minkowsky metric, constituting the unity diagonal matrix.
Hereinafter, the indices $\mu ,\nu ,\gamma $ acquire values 0, 1, 2, 3.
Indices encountered twice imply summation thereupon. Let us select harmonic
coordinates (the condition of harmonically coordinates mean selection of
concomitant frame $\frac{\partial h_{n}^{m}}{\partial x^{m}}=\frac{1}{2}%
\frac{\partial h_{m}^{m}}{\partial x^{n}}$) and let us take into
consideration that $h_{\mu \nu }$ satisfies the gravitational field equations

\begin{center}
\bigskip $\square h_{mn}=-16\pi GS_{mn}$,
\end{center}

which follow from the General Theory of Relativity; here $S_{mn}$ is
energy-momentum tensor of gravitational field sources with d'Alemberian $%
\square $ and gravity constant $G$. Then, the solution shall acquire the form

\begin{center}
$h_{\mu \nu }=e_{\mu \nu }\exp (ik_{\gamma }x^{\gamma })+e_{\mu \nu }^{\ast
}\exp (ik_{\gamma }x^{\gamma })$,
\end{center}

where the value $h_{\mu \nu }$\ is called metric perturbation, $e_{\mu \nu }$%
\ polarization, and $k_{\gamma }$\ is 4-dimensional wave vector. We shall
assume that , which constitute metric perturbation $h_{\mu \nu }$, are
distributed in space with an unknown distribution function $\rho =\rho
(h_{\mu \nu })$.

Relative oscillations $\ell $ of two particles in classic gravitational
fields are described in the General Theory of Relativity by deviation
equations

\begin{center}
$\frac{D^{2}}{D\tau ^{2}}\ell ^{i}=R_{kmn}^{i}\ell ^{m}\frac{dx^{k}}{d\tau }%
\frac{dx^{n}}{d\tau }$,
\end{center}

being $R_{kmn}^{i}$ gravitational field Riemann's tensor.

Specifically, the deviation equations give the equations for two particles
oscillations

\begin{center}
$\stackrel{..}{\ell }^{1}+c^{2}R_{010}^{1}\ell ^{1}=0,\quad \omega =c\sqrt{%
R_{010}^{1}}$.
\end{center}

The solution of this equation has the form

\begin{center}
$\ell ^{1}=\ell _{0}\exp (k_{a}x^{a}+i\omega t)$,
\end{center}

being $a=1,2,3$. Each resultant gravitational field or wave with index $j$
and Riemann's tensor $R_{kmn}^{i}(j)$ shall be corresponding to the value $%
\ell ^{i}(j)$ with stochastically modulated phase $\Phi (j)=\omega (j)t$.

We assume to be dealing with a stochastic gravitational background, so we
are to define the probability in 4-dimensional Riemann space. This
definition shall satisfy the following requirements.

We shall call the stochastic curved space the 4-dimensional Riemann space
with probabilities defined therein. With this, we shall require the
following:

1. If the interval $\Delta \ell $ in the Riemann space equals zero, then the
probability of finding the particle in this interval equals unity.

2. If the interval $\Delta \ell $ in the Riemann space equals infinity, then
the probability of finding the particle in this interval equals zero.

3. The probability to find the particle in the intervals $\Delta \ell
(x_{2}^{i},x_{1}^{i})+\Delta \ell (x_{3}^{i},x_{2}^{i})\geq \Delta \ell
(x_{3}^{i},x_{1}^{i})$, equals $P_{21}+P_{32}\leq P$.

Hereinafter, we refer to the regions of 4-dimensional Riemann space, which
are bount with each other with single type intervals, that is, only with
ether time-like intervals or space-like intervals.

It should be noted that these requirements are met by intervals with the
following type of distribution: interval $\Delta \ell $ is corresponding
with the probability interval

\begin{center}
$\Delta P=\frac{1}{\sigma \sqrt{2\pi }}\exp (-\frac{\Delta \ell ^{2}}{%
2\sigma ^{2}})$.
\end{center}

Similarly, for the probability interval of finding a particle with velocity $%
u^{i}$\ we have:

\begin{center}
\bigskip $\Delta P=\frac{1}{\sigma \sqrt{2\pi }}\exp (-\frac{S}{S_{0}})$,
\end{center}

or

\begin{center}
$\Delta P=\frac{1}{\sigma \sqrt{2\pi }}\exp (-\frac{W}{m\sigma ^{2}})$ ,
\end{center}

being $W$ the energy of oscillation of the micro-particle in question
acquired in the gravitational background; for a free particle with the RMS
velocity deviation $(\Delta u^{i})^{2}$\ this energy equals action function $%
S$. Assuming the dispersion to be characteristic for the gravitational
background and formally equating the dispersion with gravitational
background energy, we obtain

\begin{center}
$\Delta P=-\frac{1}{\sigma \sqrt{2\pi }}\exp (-\frac{S}{S_{0}})$.
\end{center}

Then, introducing notations $a^{2}=\frac{1}{\sigma \sqrt{2\pi }}$, $S_{0}=%
\frac{m\sigma ^{2}}{2}$ and taking into account that the probability is
equal to squared probability amplitude, the probability amplitude,
characterizing the action acquired by any micro-particle of mass $m$\ in the
gravitational background, will equal

\begin{center}
$\psi =a\exp (i\frac{S}{S_{0}})$.
\end{center}

The probability amplitude is a vector in the complex space. Therefore, it
can be decomposed in vector basis $e^{m}$, so that

\begin{center}
$\psi =e^{m}\psi _{m}$.
\end{center}

The scalar product of two vectors $A^{m}$ and $B^{n}$in the Riemann space is
defined as $g_{mn}A^{m}B^{n}$ , hence we can normalize this wave function as
follows:

\begin{center}
$\int g_{mn}\psi _{m}\psi _{n}dx=1$.
\end{center}

If we pass to the 3-dimensional space, having added the classical Jacobi -
Hamilton equation

\begin{center}
\bigskip $\frac{\partial S}{dt}+\frac{1}{2m}(\nabla S)^{2}+U=0$
\end{center}

and continuity equation for probability density $\psi ^{2}=a^{2}$,

\begin{center}
$\frac{\partial a^{2}}{\partial t}+div(a^{2}\frac{\nabla S}{m})=0$,
\end{center}

we shall get the Schroedinger equation for this wave function

\begin{center}
$i2S_{0}\frac{\partial \psi }{\partial t}=-\frac{4S_{0}^{2}}{2m}%
\bigtriangleup \psi +U(x,y,z)\psi $.
\end{center}

We consider the physical model with the gravitational noise [i.e. with the
background of gravitational fields and waves]. This means that we assume
existence of fluctuations in gravitational waves and fields expressed
mathematically by metric fluctuations.

Let us consider now the experiment with interference of two electrons on two
slits. The electron interference experiment is the one in which it is
impossible to determine the electron trajectory. Any attempt to determine
the electron trajectory fails due to any infinitesimal affecting of an
electron with the purpose of determination of its trajectory would alter the
interference pattern. This is the first aspect. On the other hand,
interaction of classical and stochastic fields and waves in such experiments
is usually neglected. Such interactions must exist in compliance with the
existing provisions of classical physics, and in particular, of the General
Relativity Theory. Moreover, this is experimentally confirmed by the
pre-quantum classical physics, hence, they require verification of their
effect onto quantum micro-objects.

Let us review some provisions of the General Relativity Theory. We consider
the motion of electrons from the source $S$ to the screen through slits 1
and 2.

\begin{center}
$\left\langle x\mid s\right\rangle _{1}=g_{\alpha \alpha }(1)\left\langle
x\mid s\right\rangle _{\alpha \alpha }=\delta _{\alpha \alpha }\left\langle
x\mid s\right\rangle _{\alpha \alpha }+h_{\alpha \alpha }(1)\left\langle
x\mid s\right\rangle _{\alpha \alpha }$,

$\left\langle x\mid s\right\rangle _{2}=g_{\alpha \alpha }(2)\left\langle
x\mid s\right\rangle _{\alpha \alpha }=\delta _{\alpha \alpha }\left\langle
x\mid s\right\rangle _{\alpha \alpha }+h_{\alpha \alpha }(2)\left\langle
x\mid s\right\rangle _{\alpha \alpha }$,
\end{center}

where $g_{\alpha \alpha }(1)\neq g_{\alpha \alpha }(2)$. Due to the
propagation difference between the two trajectories in space and time, the
interference pattern is generated. In the stochastic curved space one needs
not to know the electron trajectory. The interference pattern emerges due to
the difference in metrics $g_{\alpha \alpha }(1)$ and $g_{\alpha \alpha }(2)$

In this experiment we observe not interference of electrons, as usually it
is supposed, and, from the point of view of the Metrics Fluctuations Theory,
it is interference of the metrics fluctuation which are moving through two
hole. Here, electrons play the role of a medium(environment). We should add
in this process resonant properties of environment(medium), i.e. electrons.

\begin{center}
Conclusion
\end{center}

Presently, there is very little reliable information about random
gravitational background. The latter is not investigated either
theoretically or experimentally[1-9]. However, one can expect that the due
account of it is capable of providing answers for a number of questions
about the nature of microobjects.

\begin{center}
Asknowledgments
\end{center}

I'm thanks to Doctor Yuriy Svirko from Tokyo University, Department of
Applied Physics; to Professor Andrew Whitaker from Queen's Belfast
University, Physics Department and to Professor Yuiry Rybakov from People's
Friendship University of Russia, department of Theoretical Physics for help
and assistant.

\end{document}